\documentclass[]{article}

\usepackage{url}
\usepackage{appendix}
\usepackage{amsmath}
\usepackage{authblk}
\usepackage{graphicx}
\usepackage[margin=1in]{geometry}
\usepackage{url} 

\begin{document}

\title{Analog forecasting of extreme-causing weather patterns using deep learning}

\author[1]{Ashesh Chattopadhyay}
\author[1]{Ebrahim Nabizadeh}
\author[1,2,*]{Pedram Hassanzadeh}

\affil[1]{Department of Mechanical Engineering, Rice University}
\affil[2]{Department of Earth, Environmental and Planetary Sciences, Rice University}

\affil[*]{Corresponding author: pedram@rice.edu}

\maketitle

\begin{abstract}
Numerical weather prediction (NWP) models require ever-growing computing time and resources, but still, have sometimes difficulties with predicting weather extremes. We introduce a data-driven framework that is based on analog forecasting (prediction using past similar patterns), and employs a novel deep learning pattern-recognition technique (capsule neural networks, CapsNets) and an impact-based auto-labeling strategy. Using data from a large-ensemble fully coupled Earth system model, CapsNets are trained on mid-tropospheric large-scale circulation patterns (Z500) labeled $0-4$ depending on the existence and geographical region of surface temperature extremes over North America several days ahead. The trained networks predict the occurrence/region of cold or heat waves, only using Z500, with accuracies (recalls) of $69\%-45\%$ ($77\%-48\%$) or $62\%-41\%$ ($73\%-47\%$) $1-5$ days ahead. Using both surface temperature and Z500, accuracies (recalls) with CapsNets increase to $\sim 80\%$ ($88\%$). In both cases, CapsNets outperform simpler techniques such as convolutional neural networks and logistic regression, and their accuracy is least affected as the size of the training set is reduced. The results show the promises of multi-variate data-driven frameworks for accurate and fast extreme weather predictions, which can potentially augment NWP efforts in providing early warnings.
\end{abstract}

\section{Introduction}
Predicting extreme weather events such as heat waves and cold spells is of significant scientific and societal importance. However, despite decades of progress in weather prediction, mostly through improving computationally-demanding numerical weather prediction (NWP) models and data-assimilation techniques \cite{bauer2015quiet,alley2019advances}, forecasting the anomalous atmospheric circulation patterns that often drive these extreme events has remained a challenge. For example, blocking events, which are large-scale, persistent, high-pressure systems that block/divert the usual eastward winds \cite{woollings2018blocking}, have caused some of the most devastating natural disasters in recent times such as the 2003 and 2010 heat waves in Europe \cite{barriopedro2011hot,woollings2018blocking}. Yet, the state-of-the-art NWP models have difficulties with accurately predicting the formation and persistence of blocking events \cite{ferranti2015flow,matsueda2011predictability,pelly2003well}. Overall, the key characteristics of extreme-causing weather patterns, their dynamics, and conditions that favor their formation (i.e., precursors) are not well understood \cite{woollings2018blocking,teng2013probability,hassanzadeh2014responses,coumou2015weakening,horton2015contribution,hassanzadeh2015blocking,mckinnon2016long,nakamura2018atmospheric,blocksize}.

Recent advances in artificial intelligence have revolutionized how problems in various domains of business and science are approached \cite{lecun2015deep,goodfellow2016deep}. For example, in climate science, using machine learning techniques to accurately and efficiently represent unresolved physical processes in the atmosphere and ocean has produced promising results \cite{rasp2018deep,brenowitz2018prognostic,o2018using,bolton2019applications,salehipour2019deep}, and has the potential to significantly improve climate modeling and long-term climate projections in the coming years {\cite{schneider2017earth,gentine2018could,reichstein2019deep,chatto2019data}}. Moreover, deep learning techniques have been very successful in predicting some types of sequential data \cite{goodfellow2016deep}. Consequently, whether such techniques can be used for data-driven forecasting of the spatio-temporal evolution of the weather systems (and their extreme events), e.g., after training on {high-resolution} NWP model outputs or observational data, has become an active area of research. Recent efforts pursuing this approach, which essentially requires a neural network to accurately, for some time, emulate the high-dimensional nonlinear dynamics governing the evolution of the turbulent atmospheric circulation, have shown the promises and challenges of this approach {\cite{dueben2018challenges,vlachas2018data,scher2018toward,weyncan,scher2019weather,chatto2019data}}. 

In the current study, for data-driven prediction of extreme-causing weather patterns, we introduce an alternative framework that is based on analog forecasting, i.e., making prediction by finding similar pattern(s), or analog(s), in the past \cite{lorenz1969atmospheric,van2007empirical}. In the historical context, before the advent of powerful electronic computers and stable numerical schemes for integrating the partial differential equations of the NWP models, analog forecasting was a key tool in weather prediction; e.g., it was used in the planning of the D-Day for the $1944$ Normandy invasion \cite{mcdermott2016model}. Analog forecasting was used less frequently in later decades, due to the challenges in finding useful analogs and the rapid growth of NWP \cite{van2007empirical}, although the approach has the potential for a comeback given the rapid increase in data and emergence of new auxiliary methods \cite{mcdermott2016model,zhao2016analog}. 


Here, we build our data-driven framework on analog forecasting because the patterns of the circulation, e.g., the relative positions of high- and low-pressure systems, play a key role in the spatio-temporal evolution of the circulation and the initiation of extreme events at the surface, and analog forecasting essentially casts weather prediction as a complex pattern-recognition problem, an area that has been truly revolutionized by deep learning in recent years \cite{lecun2015deep,goodfellow2016deep}. Rather than looking for one perfect analog or a combination of near-perfect analogs {that are identified for example based on pattern correlation or Euclidean distance in the grid space,} as pursued in traditional analog forecasting {\cite{van1994searching,van2007empirical}}, our framework employs deep learning techniques to {classify the patterns based on their key, likely a low-dimensional, set of features and} decipher the complex relationship between this {feature space} (at the altitude of $\sim 5$km) and the extreme events (at the surface) among all training samples. The purpose of this paper is to provide a proof-of-concept for this framework.  

The structure of the paper is as follows. In Section~\ref{sec:data}, the data and definitions of extreme events and their onsets are presented. In Section~\ref{sec:method}, the data-driven extreme weather prediction framework, including the labeling, training, and testing procedures, are discussed. Results are presented in Section~\ref{sec:results} followed by Discussion in Section~\ref{sec:discussion}.

\section{Data}\label{sec:data}
\subsection{LENS Data} \label{sec:data}
We use daily data from the Large-Ensemble (LENS) Community Project \cite{kay2015community}, which consists of a $40$-member ensemble of fully-coupled atmosphere-ocean-land-ice Community
Earth System Model version 1 (CESM1) simulations with $1920-2005$ historical radiative forcing. For each ensemble member, the same historical radiative forcing is used, but random, weak perturbations are added to the initial state of each member to create an ensemble. To ensure abundant training samples for the purpose of demonstrating a proof-of-concept for the framework, we choose to use data from a large-ensemble climate model, rather than reanalysis data (see Discussion). Still, the simulated atmospheric circulation is non-stationary, turbulent, and multi-scale, with complexities similar to those of the real atmosphere, thus providing a challenging testbed for our data-driven extreme weather prediction framework. 

From this dataset, we use surface air temperature, measured as temperature at 2m above ground (T2m), and geopotential height at 500mb (Z500). We use the daily-averaged T2m and Z500 from 1920-2005 for the months of June-August (boreal summer) and December-February (boreal winter) from all 40 ensemble members.

\subsection{Extreme hot and cold events and their onsets} \label{sec:onset}
We focus on extreme temperature events over the North American continent in the subtropical and midlatitude regions between $30^\mathrm{o}$N and $60^\mathrm{o}$N. For a given calendar date in a given ensemble member, the T2m anomalies are computed by removing the climatological
mean, defined as the $15$-day running mean of T2m centered around that date and averaged over all ensemble members. Following Chan {\it et al.}\cite{chan2019evaluating}, heat waves (cold spells) are defined as land grid points over North America in summer (winter) with daily T2m anomaly in the $99$ ($1$) percentile and larger (smaller) than $3$K ($-1$K) for a sequence of at least $5$ consecutive days. We then identify the onsets of these extreme temperature events as the first day of each sequence.


Unlike data commonly used in deep learning applications, climate data have spatio-temporal correlations, which can affect the training process (e.g., by reducing the effective sample size) and/or can lead to artificially high accuracies during testing (e.g., if the training and testing sets have strongly correlated samples). Here we aim to choose distinct samples of onsets with minimal correlations within and across the training and testing sets. In constructing the dataset onsets to be used for training/testing, we follow these criteria: We include an onset if there is no other extreme events within the earlier $5$ days and if it is not followed by another extreme event within the next 10 days (i.e., maximum of one onset in each $16$-day window over North America). This procedure substantially removes potentially (temporally) correlated Z500 patterns corresponding to persistent heat waves/cold spells that may artificially enhance prediction skills. We have experimented with the size of this window ranging from $16-24$ and found no noticeable changes in the reported accuracies. For non-extreme events, we ensure that their T2m anomaly is not in the $99$ $(1)$ percentile and that they are not chosen from the events in the $16$-day windows defined above. 

{As discussed later, we randomly divide the samples in this dataset of onsets (generated using all $40$ ensemble members) to a training set and a testing set. To absolutely ensure that the reported accuracies are not contaminated by any temporal correlation between the samples in the training and testing sets (even after the procedure described above), we conduct another analysis in which data from $30$ ensemble members are used in the training set and data from the $10$ remaining ensemble members are used in the testing set (thus, absolutely no temporal correlation between the training and testing samples).  The analysis demonstrates the same accuracies as those reported in the paper.}



\subsection{Geographical clustering of the extreme events' onsets}
\begin{figure}[ht]
  \centering
  \includegraphics[width=1\textwidth]{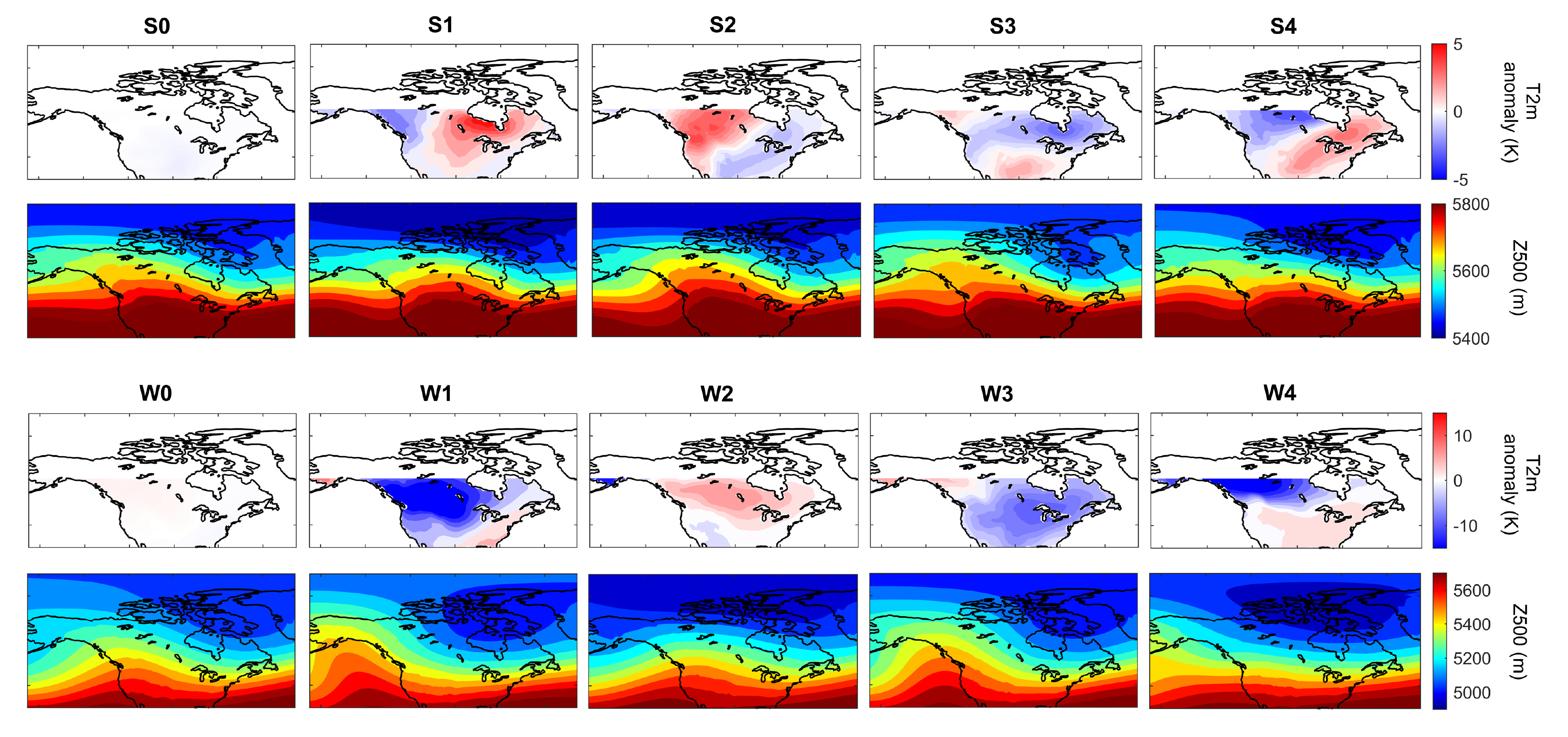}
  \caption{Cluster centers of T2m anomalies at the onsets and Z500 patterns of 3 days earlier. The top (bottom) two rows correspond to summers (winters). S0 (W0) shows the average of T2m and Z500 patterns from days with no heat wave (cold spell). S1-S4 and W1-W4 are obtained from K-means clustering the anomalous T2m patterns at onsets into four classes, which roughly separates the extreme events into four geographical regions: Northern Canada (S1), Western US-Canada (S2), Southern US (S3), and Eastern US-Canada (S4) in summers, and North-West US-Canada (W1), Alaska (W2), North-East US-Canada (W3), and Northern Canada (W4) in winters. Rows 1 and 3 show the cluster centers while rows 2 and 4 show the average of Z500 patterns 3 days before the onsets for each cluster.}
  \label{fig:Clusters}
\end{figure}

We cluster the onsets of extreme events into four distinct yet cohesive geographical regions, separately for winters and for summers. Following Viguad {\it et al.}\cite{vigaud2018multiscale}, an empirical orthogonal function analysis is first performed on the T2m patterns on the onset days and the first $22$ principal components (PCs), which explain over $90\%$ of the variance, are retained. The K-means algorithm \cite{lloyd1982least} is then used on the retained PCs and repeated $1000$ times with new initial cluster centroid positions, and a cluster index $1$, $2$, $3$ or $4$ is assigned to each day. Rows 1 and 3 of Figure~\ref{fig:Clusters} show the cluster centers (in terms of T2m). 

{The aim of our data-driven framework is to predict whether in a few days, a Z500 pattern would lead to no extreme event (assigned cluster $0$) or an extreme event in the geographical regions corresponding to clusters $1-4$}. Note that here we focused on four clusters to have a balance between too many clusters that may not be distinct, or too few that would not effectively separate the geographical regions, and demonstrate the effectiveness of the proposed data-driven extreme weather event prediction framework. Also, here we use K-means, a simple unsupervised algorithm, but other methods such as hierarchical clustering or self-organizing maps \cite{horton2015contribution, mo1988cluster, cheng1993cluster} could be used instead. 


\section{Methodology}\label{sec:method}
\subsection{Deep Learning Techniques: ConvNet and CapsNet}
We use two state-of-the-art deep learning techniques for pattern recognition: Convolutional neural network (ConvNet or CNN) \cite{lecun2015deep,goodfellow2016deep} and a more advanced method, capsule neural network (CapsNet) \cite{sabour2017dynamic}. The key advantage of both methods over traditional image-processing techniques is that the filters used for feature extraction are learned for each dataset through an algorithm called backpropagation \cite{goodfellow2016deep}, rather than being hand-engineered and specified beforehand. ConvNet is the groundbreaking method that has transformed image processing since $2011$, but because of a property called equivariance that is discussed later, CapsNet is expected to work even better for our {spatio-temporal climate data}. Figure~\ref{fig:Cartoon} shows the architecture of the CapsNet schematically. Details of the ConvNet and CapsNet architectures are presented in \ref{cnn} and \ref{capsnet}, respectively.     


\subsection{Impact-based auto-labeling of daily Z500 patterns} \label{sec:label}
Both ConvNet and CapsNet are supervised methods, meaning that they have to be first trained on labeled patterns. However, given the incomplete understanding of extreme-causing weather patterns and their complexities, expert-labeled data {are} not useful for our objective. For example, indices designed to find blocking patterns in the Z500 field based on their presumed properties are known to perform poorly, e.g., in identifying extreme-causing patterns even on the same day as the heat or cold extreme events \cite{chan2019evaluating}. Expert-labeling becomes even less effective for the purpose of prediction, which requires accounting for the nonlinear spatio-temporal evolution of the atmospheric circulation over several days or longer. 

To overcome this challenge, here we devise an impact-based auto-labeling strategy: {Knowing} the surface temperature over North America on a given day, the Z500 pattern of several days earlier is labeled as $0$ (no extreme onset) or $1$, $2$, $3$ or $4$ (the cluster indices of T2m extremes). For example, for predicting heat waves in summers with a lead time of 3 days, the Z500 patterns 3 days before the onsets are labeled S0-S4 based on the cluster index of the T2m at the onsets. Rows 2 and 4 of Figure~\ref{fig:Clusters} show the average of Z500 patterns with the same labels for $3$-day prediction. For the Z500 patterns, we use two domains with the same latitudinal ($30^\circ$N-$90^\circ$N) but different longitudinal extends: $195^\circ$E-$315^\circ$E (small domain, shown in Figures~\ref{fig:Cartoon}-\ref{fig:Clusters}) and $145^\circ$E-$340^\circ$E (large domain). For winters, at prediction lead time beyond $2$ days, we have found higher accuracy and recall with the large domain, which is likely due to the dominance of zonal advection and Rossby wave propagation in the evolution of weather patterns. For summers, we have found higher accuracy with the small domain for all prediction lead times. The results in Figures~\ref{fig:Fig3}-\ref{fig:Fig6} are with large (small) domain for winters (summers). 

We highlight that in conventional deep learning applications, labeling and training/testing are all conducted on the same feature map; however, in the impact-based labeling strategy introduced here, we label based on differences in one feature map (T2m) but train/test on another feature map (Z500), in order to predict the original feature map (T2m). While more challenging, the impact-based auto-labeling strategy circumvents the need for a full understanding of the complex and nonlinear relationship between the predictor (Z500) and the impact of interest (T2m).

\subsection{Training and testing datasets} 
For each season and prediction lead time, we build a dataset of $M$ labeled Z500 patterns per cluster. While the number of onsets vary among the clusters and there are many more non-extreme events than extreme events, to avoid \textit{class imbalance}, we choose $M$ near the smallest number of onsets among the four clusters of extreme events. For both summer and winter, we use $M=1000$ for prediction lead times of $1$ and $2$ days, and $M=900$ for longer lead times. For each pair of training and testing sets, we randomly choose $N=3M/4$ samples per cluster for the training set and the remaining $M/4$ samples for the testing set. Note that the total number of samples in the training and testing sets ($\sim 4500-5000$) is much lower than the total number of summer or winter days ($316480$ and $306000$, respectively) in the LENS dataset, because we are focusing on the rare events (i.e., onsets of extremes in the $99$ or $1$ percentile). We also report the accuracies with smaller training sets of sizes $3N/4$, $N/2$, and $N/4$ samples per cluster in Figures~\ref{fig:Fig3}-\ref{fig:Fig6} (the number of samples in the testing set is kept as $M/4$ regardless of the size of the training set).  

\subsection{Data-driven extreme weather prediction framework}
\begin{figure}[ht]
  \centering
  \includegraphics[width=1\textwidth]{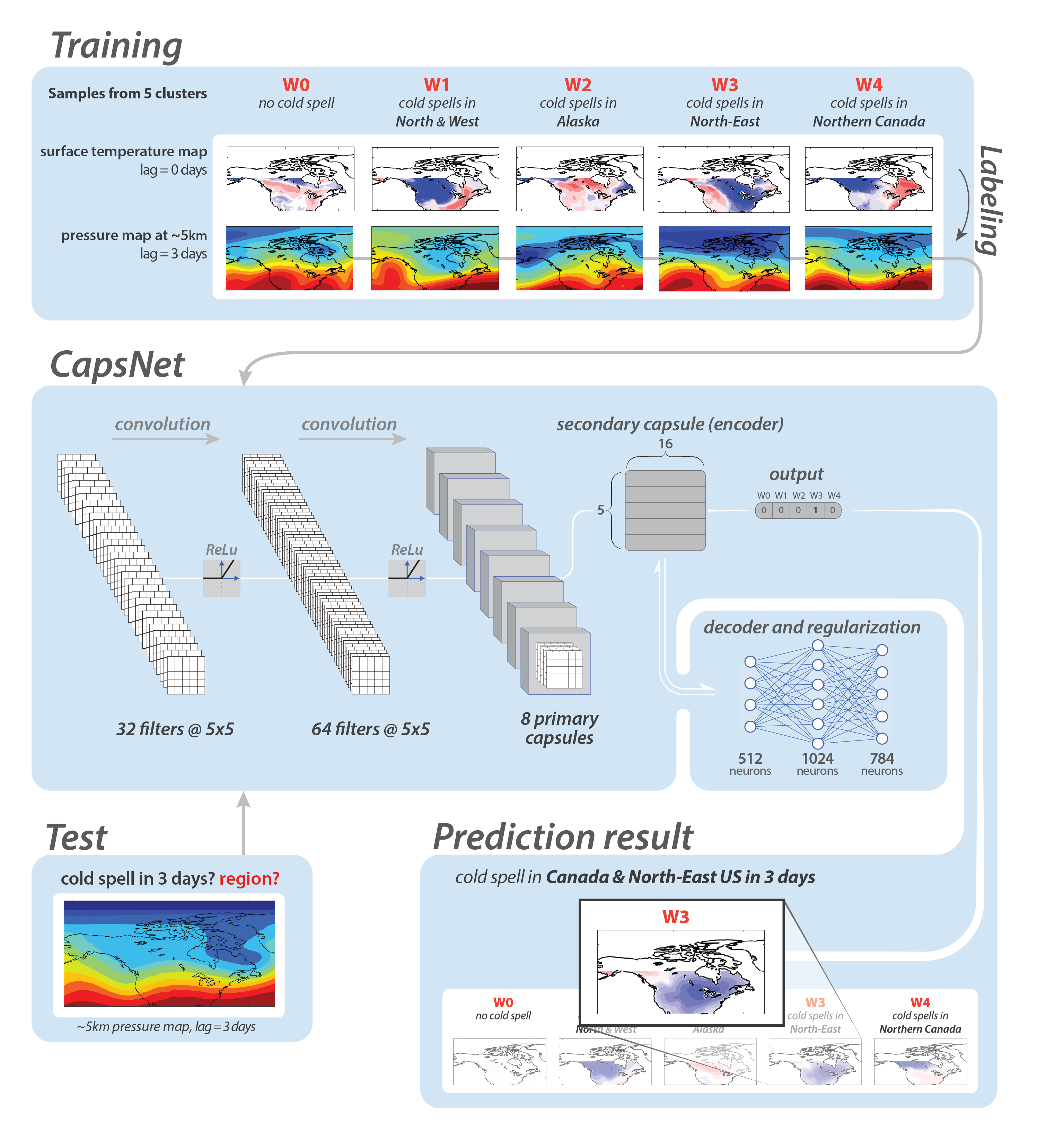}
  \caption{Schematic of the data-driven framework for prediction of cold spells based on Z500 patterns of $3$ days earlier. Using the impact-based auto-labeling strategy, Z500 patterns are labeled W0, W1, W2, W3, or W4, depending on the cluster index of T2m three days ahead. The panels at the top show examples of T2m patterns at the onset and the corresponding Z500 patterns (from three days earlier) for each cluster. Only the Z500 patterns and their labels are inputted into the CapsNet during training. Once trained, the CapsNet can predict, from a given Z500 pattern, the T2m cluster index of three days later, thus predicting the occurrence and geographical region of cold spells. For the shown test example, a cold spell in W3 in $3$ days is predicted. Note that for winters, Z500 patterns over a larger domain that extends across the Pacific Ocean to $145^\circ$E are inputted into the CapsNets, but a smaller domain is shown in this figure for better illustration (see Section~\ref{sec:label}). Separate CapsNets are trained for each season and each prediction lead time.}
  \label{fig:Cartoon}
\end{figure}

     The schematic of the entire data-driven prediction framework is shown in Figure~\ref{fig:Cartoon}. Separate CapsNets (or ConvNets) are trained and used for different seasons and prediction lead times (referred to as Cases hereafter). For example, Figure~\ref{fig:Cartoon} shows the framework with CapsNet for prediction of cold spells in winter with a $3$-day lead time. During the training phase, the same number of Z500 patterns ($N$ per cluster) and their cluster indices from the training set are inputted into the CapsNet. The trained CapsNet can then predict the cluster index of a never-seen-before Z500 pattern inputted from the testing set. If the output of CapsNet is index W0, then no cold spell anywhere over North America (between $30^\circ$N-$60^\circ$N) is forecasted in $3$ days, while other outputs indicate prediction of a cold spell in $3$ days over North-West US-Canada (W1), Alaska (W2), North-East US-Canada (W3), or Northern Canada (W4).

     The above framework for predicting the extreme-causing weather patterns (just based on information from Z500) is used for the results of Figures~\ref{fig:Fig3} and \ref{fig:Fig5}. To further demonstrate the potentials of this data-driven framework in predicting the extreme events in a multi-variate approach, we have also shown results in Figures~\ref{fig:Fig4} and \ref{fig:Fig6} with an extended framework, in which the inputs consist of the patterns of both Z500 and anomalous T2m stacked together in two different channels. Due to the difference in their mean and variability, standardization is performed on each channel separately.   

     
\subsection{Training and testing procedure and accuracy/recall calculations}
For each Case, the $M$ labeled samples are randomly divided into a training set and testing set (with ratio of $3:1$, as discussed above) four times to create four pairs of training/testing sets. One pair is used as a validation set to explore the CapsNet and ConvNets' hyperparameters such as kernel size, regularization constant, and dropout probability. For each Case, once a suitable set of hyperparameters is found, the CapsNet (or ConvNet) is trained and tested just once on each of the remaining three pairs of the training/testing sets (to emphasize, the hyperparameters are not changed for these three pairs, but the weights of the filters are learned independently each time). The accuracies and recalls reported in Figures~\ref{fig:Fig3}-\ref{fig:Fig6} and in the text are computed as the average of results with these three pairs of datasets. We have adopted this approach to examine the robustness of the results despite the relatively small size of the labeled dataset.      

We report the prediction skills in terms of the total accuracy of the testing set, computed as the number of test samples from all $5$ clusters whose cluster index is correctly predicted divided by the total number of test samples, and recall, computed as the number of test samples from the four clusters with extreme events $(1-4)$ whose cluster index is correctly predicted divided by the total number of test samples in clusters $1-4$. We computed the recall because for extreme weather prediction, missed events are much more undesirable than false alarms. Together, accuracy and recall fully quantify the skills of the framework for a multi-class prediction. {Note that the} accuracy for individual clusters, computed as the number of correctly predicted test samples from that cluster divided by the total number of test samples from that cluster, is the receiver operating characteristic (ROC) score, a common forecast skill metric \cite{fawcett2006introduction,mckinnon2016long}. 

\section{Results}\label{sec:results}
\label{results}
\begin{figure}[ht]
  \centering
  \includegraphics[width=1\textwidth]{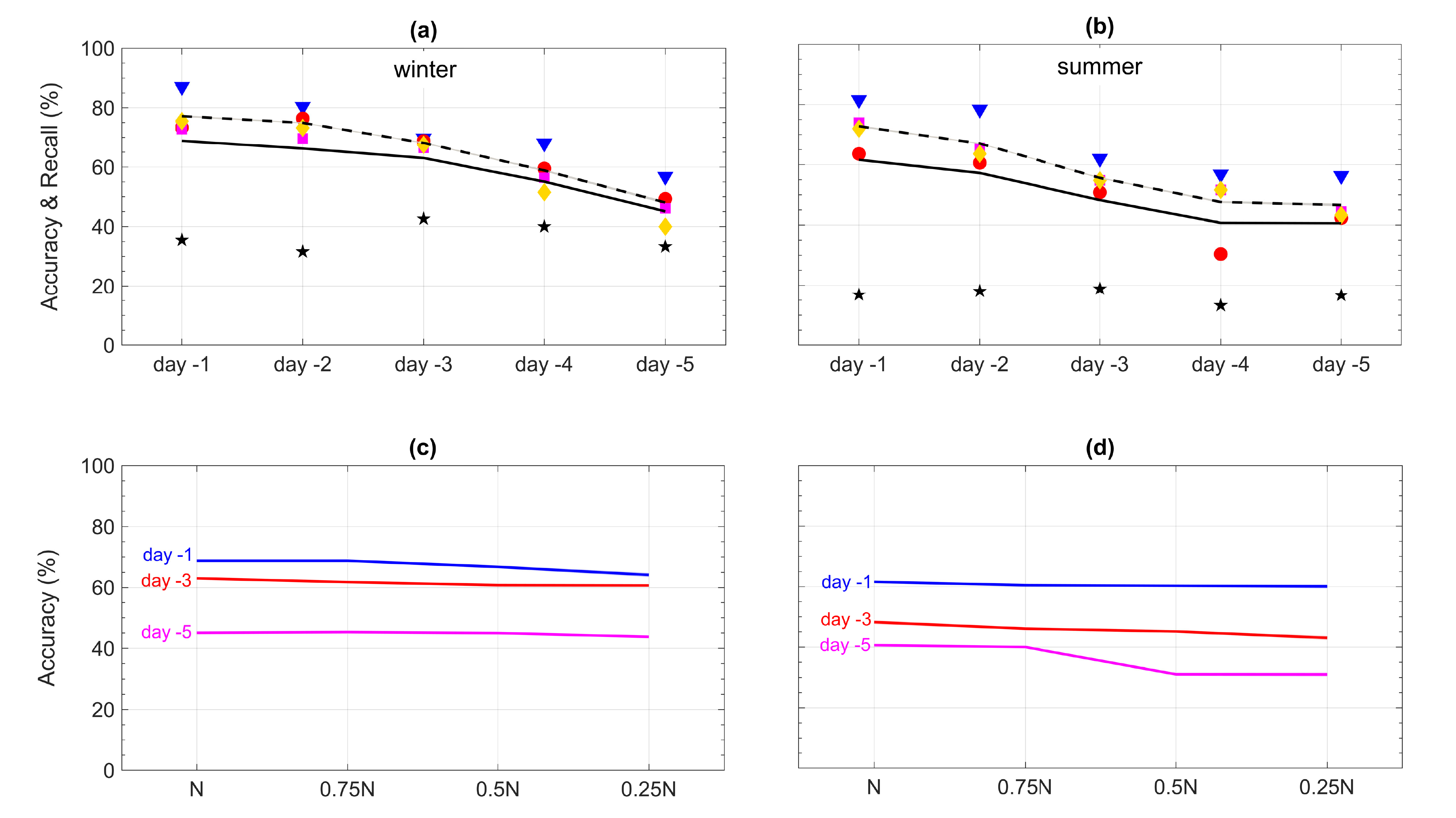}
  \caption{Performance of CapsNets in predicting heat waves and cold spells using Z500 patterns at various lead times. (a)-(b) The symbols show the accuracy at different lead times for each cluster: star (0), triangle (1), square (2), diamond (3), and circle (4). The solid (dashed) lines show the total accuracy (recall). (c)-(d) Total accuracy at prediction lead times $1$, $3$, and $5$ days versus the size of the training set ($N=750$; fractions are rounded to the nearest integer if needed). Results in (a)-(b) are obtained with the largest training set. The symbols show the accuracy averaged over $3$ randomly-drawn pairs of training/testing sets. The lines and their shading depict the mean and $\pm{1}$SD of accuracy or recall computed for the 3 pairs; the shadings are narrow, demonstrating the small SD and robustness of the results.}
  \label{fig:Fig3}
\end{figure}

Figure~\ref{fig:Fig3} shows the performance of CapsNet for predicting cold spells and heat waves using the Z500 patterns from $1-5$ days earlier. The accuracies for lead times of $1$ to $5$ days are between $68.8\%\pm{0.3\%}$ and $45.1\%\pm{0.1\%}$ in winter, and $61.6\%\pm{0.0\%}$ and $40.6\%\pm{0.1\%}$ in summer, against a $20\%$ random chance in a 5-class prediction. The recalls are consistently higher, between $77.2\%\pm{0.3\%}$ and $48.1\%\pm{0.1\%}$ in winter and $72.8\%\pm{0.1\%}$ and $46.6\%\pm{0.1\%}$ in summer. Examining the prediction accuracy for individual clusters shows that the inaccuracies largely result from false alarms due to non-extreme-events (cluster 0) incorrectly predicted as an extreme event somewhere in North America (clusters $1-4$). False alarms can be reduced by adding more constraints on Z500 during labeling, e.g., requiring daily Z500 anomalies to exceed $1.5$ standard deviation (SD); however, we {choose} to avoid subjective criteria and {only use} the impact (i.e., T2m extreme) for labeling. Furthermore, we {focus} on minimally pre-processed inputs, e.g., we do not de-trend Z500 patterns and instead use the full Z500 patterns (see Section~\ref{sec:label}), which are non-stationary due to low-frequency coupled atmosphere-ocean modes of climate variability and changes in the radiative forcing from $1920-2005$. 

     The results in Figures~\ref{fig:Fig3}(a)-(b) are obtained with a training set containing $N=750$ samples from each of the $5$ clusters. Figures~\ref{fig:Fig3}(c)-(d) show that as the size of the training set is reduced, the accuracies for winter barely decline. Even when the number of training samples per cluster is reduced almost by a factor of $4$ to $187$ or $168$ (depending on the lag), the largest decrease in accuracy is $4.7\%$ (for day $-4$). In summer, the effect of the size of the training set is more pronounced especially at longer lead times, e.g., the accuracy for 5-day prediction declines by $9.5\%$ when the training set is reduced by a factor of $4$. Overall, the weak dependence of the accuracy on the size of the training set is encouraging for practical purposes (see Discussion), but it also suggests that likely, higher accuracies could not be achieved even if we had more training samples (see below). 
     
     The prediction skills in summers are lower than those in winters. Figure~\ref{fig:Clusters} shows that the Z500 patterns corresponding to different clusters are much more similar in summers than in winters, suggesting that it should be harder to identify the correct cluster of a pattern in summer. Still, that CapsNet can differentiate between patterns that have such similar averages (i.e., cluster centers) with the accuracy (recall) of $48.2\%\pm{0.1\%}$ $(55.6\%\pm{0.1\%})$  shows the effectiveness of the framework. Furthermore, dynamics of heat waves are more complex than cold spells and the mid-tropospheric circulation patterns (the only predictor here) are not the only driver: {Cold} spells are mostly due to equatorward advection of cold air from higher latitudes while the heat waves are caused by a combination of horizontal advection and adiabatic and clear-sky radiative warmings \cite{dole2011there, pfahl2012quantifying, schneider2015physics}. Moreover, land-atmosphere feedbacks play a role in the dynamics of heat waves \cite{miralles2014mega}.

     The results of Figure~\ref{fig:Fig3} show the power of our data-driven framework for predicting the surface temperature extreme events using a single variable (Z500) that represents mid-tropospheric circulation, i.e., predicting extreme-causing weather patterns. The above discussions on the weak dependence of accuracy on the size of the training set and the dynamics of the extreme temperature events suggest that including more variables as the predictor and pursuing a multi-variate framework would lead to better prediction skills for the extreme temperature events, particularly at longer lead times. It should be highlighted that even for winters, where meridional advection dominates, including information from other altitudes of troposphere and stratosphere (e.g., to account for polar vortex variability) are expected to improve the prediction skills (see Discussion).  

\begin{figure}[ht]
  \centering
  \includegraphics[width=1\textwidth]{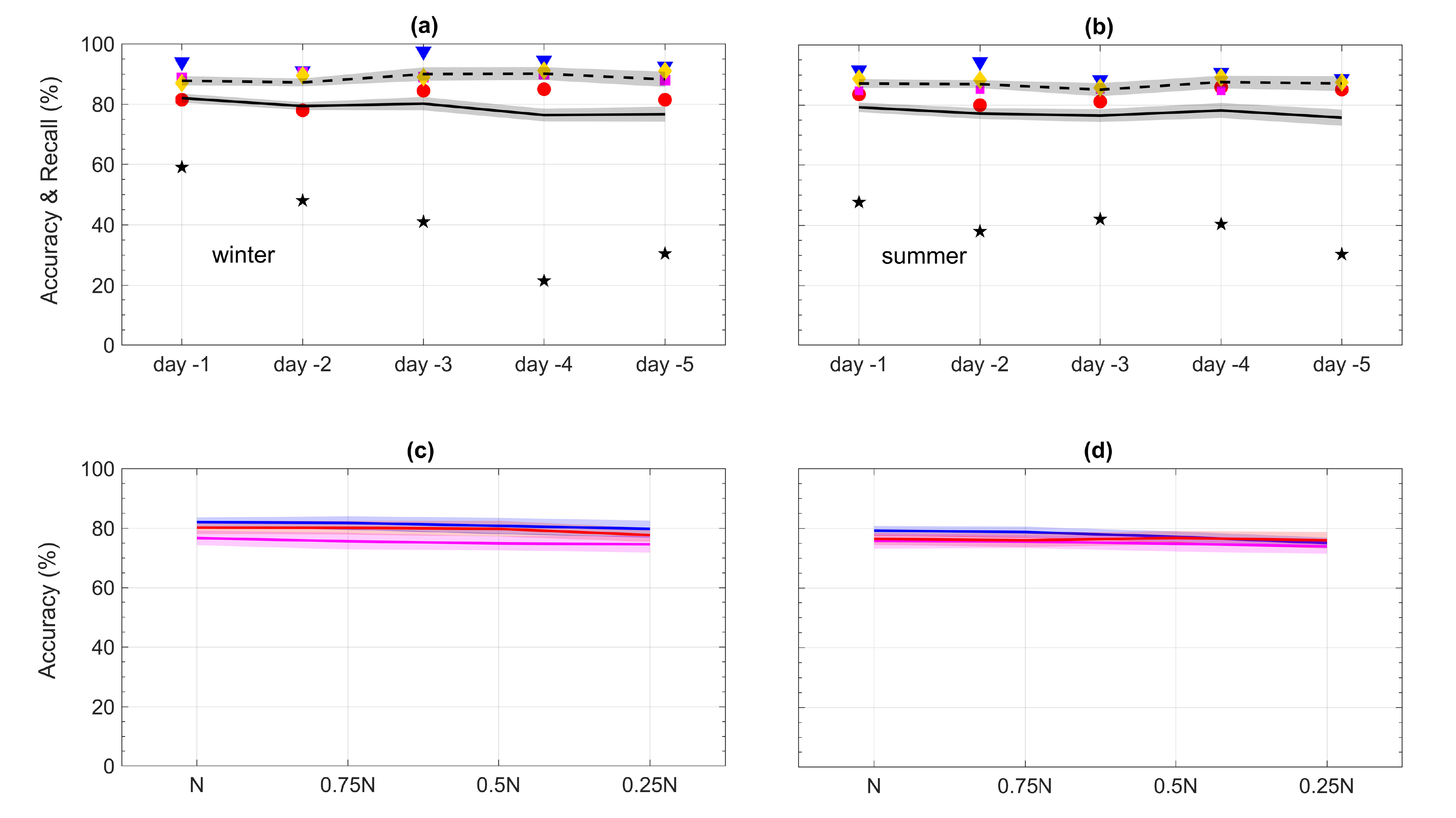}
  \caption{Same as Figure~3 but for the performance of CapsNets in predicting heat waves and cold spells using both T2m and Z500 patterns. The shadings show $\pm{1}$SD; the SD values are higher in the multi-variate approach. In (c)-(d), the change of accuracies with the size of the training set is small for lead times of $1$, $3$, and $5$ days (labels not shown).}
  \label{fig:Fig4}
\end{figure}

     To demonstrate the promises of such multi-variate data-driven frameworks, we {repeat} the analysis of Figure 3 but by inputting the patterns of Z500 and anomalous T2m together into CapsNet in the training and testing phases. Figure~\ref{fig:Fig4} shows that the accuracies (recalls) for lead times of $1$ to $5$ days rise to between $82.0\%\pm{1.5\%}$ $(87.8\%\pm{1.4\%})$ and $76.7\%\pm{2.5\%}$ $(88.2\%\pm{2.3\%})$ in winter and $79.3\%\pm{1.6\%} (87.2\%\pm{1.7\%})$ and $75.8\%\pm{2.7\%}$ $(87.2\%\pm{2.6\%})$ in summer, significantly improving the prediction skills, particularly in the longer lead times. With T2m included, the false alarms decline in most cases, and the accuracies/recalls hardly change with lead time or size of the training set. It should be highlighted that the high prediction skills with Z500+T2m are not simply due to the temporal memory as a result of including T2m of earlier days. With $N=750$, training and testing the CapsNets on T2m alone result in accuracies that are consistently lower, between $0.6\%-5.2\%$ $(1.5\%-4.5\%)$, than the accuracies with Z500+T2m in winter (summer), showing that information about the atmospheric circulation adds to the {prediction} skills.   

     The accurate and robust $1$ to $5$-day predictions in Figure~\ref{fig:Fig4} suggest that the multi-variate framework using Z500+T2m, or even more variables (see Discussion), might have high prediction skills for lead times beyond $5$ days. However, such longer predictions will require using Z500 patterns (and some of the other variables) at the planetary-scales, which, for the best performance of the framework, needs CapsNet (and ConvNet) architectures capable of accounting for the Earth's spherical geometry, e.g., the zonal periodicity and decrease of area with latitude. Extending the framework to planetary-scales and longer prediction lead times is left for future work, which will benefit from recent advances in developing spherical ConvNets \cite{cohen2018spherical,jiang2019spherical}.

\begin{figure}
\centering
\noindent\includegraphics[width=1\textwidth]{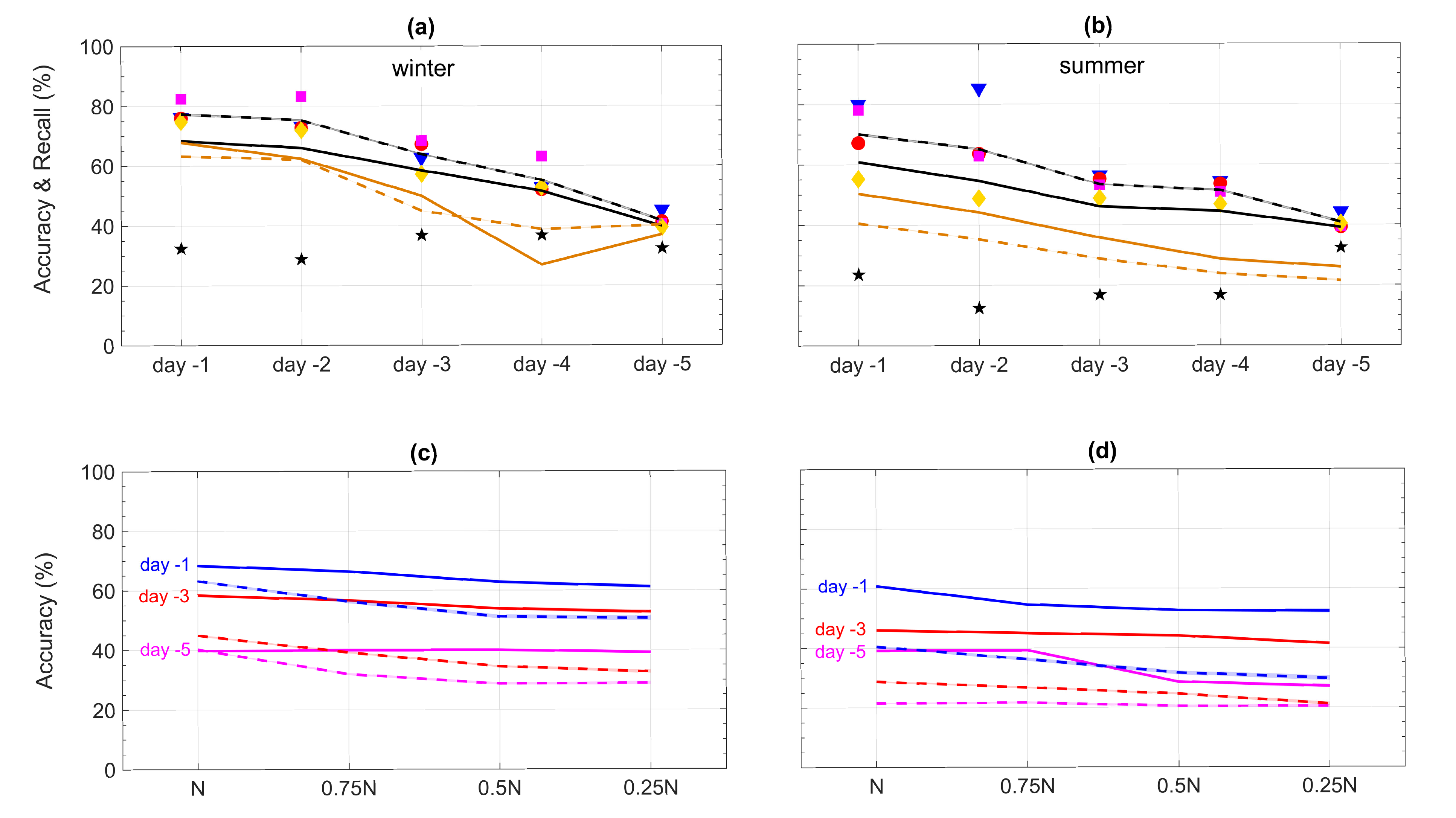}
  \caption{Same as Figure~\ref{fig:Fig3} but for the performance of ConvNets and logistic regression in predicting heat waves and cold spells using Z500 patterns at various lead times. In (a)-(b), the symbols and black lines correspond to ConvNets, while the green lines correspond to logistic regression. In (c)-(d), the solid (dashed) lines correspond to ConvNets (logistic regression). }
    \label{fig:Fig5}
\end{figure}

\begin{figure}
 \centering
\noindent\includegraphics[width=1\textwidth]{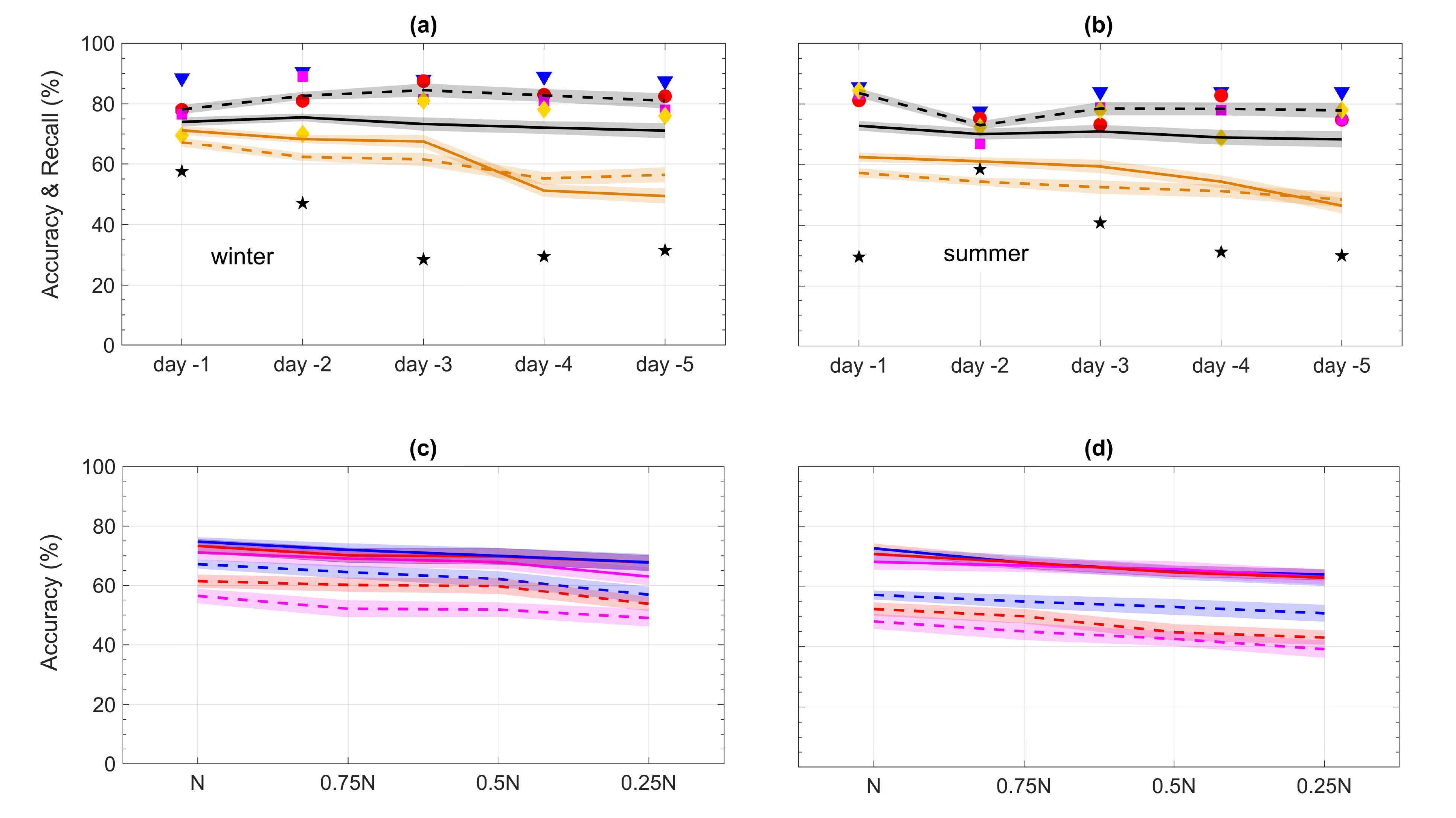}
  \caption{Same as Figure~\ref{fig:Fig4} but for the performance of ConvNets and logistic regression in predicting heat waves and cold spells using both T2m and Z500 patterns at various lead times. In (a)-(b), the symbols and black lines correspond to ConvNets, while the green lines correspond to logistic regression. In (c)-(d), the solid (dashed) lines correspond to ConvNets (logistic regression).}
    \label{fig:Fig6}
\end{figure}

     We {also conduct} the analyses in Figures~\ref{fig:Fig3} and \ref{fig:Fig4} with CapsNet replaced with two simpler methods: i) ConvNet, which is a deep learning method of growing interest in climate science (and other disciplines) and was used, for example, by Liu {\it et al.}\cite{liu2016application} to identify tropical cyclones and atmospheric rivers and by Ham {\it et al.}\cite{ham2019deep} for multi-year ENSO prediction, and ii) {Logistic} regression \cite{goodfellow2016deep}, which is a widely-used machine learning method that has been employed in some past weather forecasting efforts \cite{applequist2002comparison,herman2018money,whan2018comparing,chattopadhyay2018test}. Figures~\ref{fig:Fig5} and \ref{fig:Fig6} show that the CapsNets consistently outperform the ConvNets (except for one Case: 4-day lead time in summer). For predictions with Z500 (Z500+T2m), the accuracies of CapsNets are, on average, higher than ConvNets by $2.8\%$ $(7.7\%)$ in winters and $0.7\%$ $(7.1\%)$ in summers. {More importantly}, as the size of the training set is reduced, the accuracy of ConvNets degrades more than that of CapsNet, particularly in the multi-variate approach with Z500+T2m (Figures~\ref{fig:Fig5}-\ref{fig:Fig6}(c)-(d)). Due to their different architectures (see \ref{cnn} and \ref{capsnet}), CapsNets extract more features and information from each pattern compared to ConvNets, and are thus expected to work well even with relatively small training sets. {Moreover}, CapsNets account for the relative position and orientation of features (a property called equivariance) \cite{sabour2017dynamic}. Relative positions of features in {spatio-temporal} climate data are important, e.g., high-pressure systems on the poleward side of low-pressure systems might stall and cause weather extremes, while low-pressure systems on the poleward side of high-pressure system often move eastward without causing extreme events.

     The accuracy of logistic regression is consistently lower than that of ConvNets (and thus CapsNet), see Figures~\ref{fig:Fig5}-\ref{fig:Fig6}. For predictions with Z500, the accuracies of CapsNets are, on average, higher than those of logistic regression by $11.4\%$ $(19.6\%)$ in winters (summers). These results show the advantage of more advanced deep learning techniques over simpler ones such as ConvNet and logistic regression, and suggest that future studies in climate and environmental sciences might benefit from using CapsNets (and might benefit even more from deep learning techniques designed specifically for multi-scale, spatio-temporal, chaotic data).  
     
    Note that we {did not compare} the performance of our framework with persistence or climatology, which are two common baseline methods \cite{murphy1992climatology}, because they could not be formulated to predict T2m extremes based on inputs of Z500 patterns, and that by definition, there is no T2m extreme within $5$ days of the onsets (see \ref{appC}). 
     
     
\section{Discussion}\label{sec:discussion}
The results of Figure~\ref{fig:Fig3} show the skills of the data-driven framework in predicting high-impact events (e.g., T2m extremes) only through limited information about the events's driver (or one of the key drivers), i.e., Z500 patterns in this case, and without any information about the impact itself. This skillful prediction of extreme-causing weather patterns provides a proof-of-concept for the framework (Figure~\ref{fig:Cartoon}). We emphasize that the key components of this data-driven framework are the novel impact-based auto-labeling technique and the power of CapsNets in pattern recognition, which together enable the framework to decipher the relationship between the T2m and Z500 patterns, and the temporal evolution of Z500 patterns, despite challenges such as sensitivity of nonlinear systems to small perturbations in initial conditions \cite{lorenz1969atmospheric} {and the rarity of perfect analogs in climate data \cite{van1994searching,van2007empirical}}.


     Based on the results of Figure~\ref{fig:Fig4}, the multi-modal framework {(in which both Z500 and T2m are used together)}, once equipped with spherical CapsNets for planetary-scale inputs, may offer a promising data-driven approach to prediction. Higher accuracies and longer prediction lead times (at the weekly to seasonal time-scales, which are of the most utility and interest) might be achievable by including variables such as geopotential heights at more tropospheric levels, soil moisture, and outgoing longwave radiation, as well as information from the slow-varying boundaries of the troposphere such as tropical and extratropical sea-surface temperature (e.g., from Pacific Ocean), tropical atmosphere (e.g., Madden-Julien Oscillation), sea ice, and stratosphere, which are all known to enhance predictive skills for the midlatitude extreme events \cite{mckinnon2016long, baldwin2001stratospheric,mundhenk2018skillful}.        

     The data-driven extreme event prediction framework introduced here can be useful in (at least) two ways: 1) {To} provide early warnings of extreme events and guide the public and NWP efforts, and 2) {To} identify the precursors of extreme events using ideas from interpretable machine learning \cite{zeiler2014visualizing}. Regarding the former, one of the most appealing and powerful aspects of a data-driven framework is the possibility of training on observational data. Here, the main challenges in using observed climate data for training are that such records are short, and the data are non-stationary in time. Reanalysis products are available for as early as $1850$, although the data from before $1979$ are derived from limited direct observations. The LENS data used in this study have complexity and non-stationarity similar to that of the reanalysis data, however, the 40-member ensemble simulations provide, e.g., $\sim 300000$ days of data in winters, which is much larger than what can be obtained from reanalysis datasets. Given our focus on the onsets of extreme events in the $1$ or $99$ percentile, from the LENS data, we only used as high as $750$ and as low as $168$ samples per cluster for training. Figures~\ref{fig:Fig3} and \ref{fig:Fig4} show that even with the smallest training set, skillful multi-class predictions are obtained. Furthermore, transfer learning can be used to first train the CapsNet (or ConvNet) on a large set of modeled data and then on a small set of reanalysis, as for example, recently done by Ham {\it et al.}\cite{ham2019deep} for ENSO forecasts. The above discussion suggests that it might be possible to use data derived from observations (and not just NWP or climate models) for training of the data-driven framework. We highlight again that the purpose of this paper is to provide a proof-of-concept for the framework. Evaluating the performance of the framework trained on reanalysis data, and comparing the forecast skills with those of the NWP models, are admittedly the essential and important next steps, and are currently underway.   

     Data from large-ensemble, high-resolution NWP model simulations can also be used for training. The very high-resolution NWP models, which need prohibitive computing time/resources to run continuously, simulate the atmospheric circulation, and in particular extreme-causing patterns, with higher fidelity compared to the simulations with lower resolutions \cite{alley2019advances,jung2012high}. The advantage of using the data-driven framework, trained on high-resolution NWP models, is that it can yield extremely fast and inexpensive regional predictions, which can provide early warnings and guide the deployment of computing and sensing resources for large-ensemble, high-resolution NWP of a region predicted to experience an extreme event in several days (or longer).            

      In this study we conducted 5-class predictions based on extreme events over North America clustered (using K-means algorithm) into $4$ geographical regions. However, other clustering algorithms, number of clusters etc. could be used, or alternatively, separate data-driven frameworks can be developed for binary (yes/no) extreme predictions in each region of interest, e.g., around Greater Houston, Bay Area, Greater Boston. Understanding which of the approaches  discussed above (differing in training data and framework design) lead to the best data-driven prediction skills and better handle practical limitations requires extensive research and should be addressed in future studies.

     Precursors of extreme-causing weather patterns such as blocking events are not well understood \cite{woollings2018blocking,hassanzadeh2015blocking,nakamura2018atmospheric} and identifying them can lead to a better dynamical understanding of extreme events and potentially improving weather and climate models. Given that CapsNets show skills in predicting the extreme-causing weather patterns, it is of interest to understand how the neural network has {\emph{learned}} what key features to look for. However, understanding how deep neural networks work is known to be challenging and an active area of research. In future work, the feature maps and filters should be examined to seek an understanding of how CapsNets differentiate between patterns that in a few days lead to different T2m clusters. Recent papers by Ham {\it et al.}\cite{ham2019deep} and Mcgovern {\it et al.}cite{mcgovern2019making} present promising results and discuss interesting ideas on interpretation of ConvNets and other machine learning methods applied to climate data.      
        

     Finally, our data-driven framework can be readily generalized for prediction of other high-impact climatic and environmental phenomena, e.g., extreme precipitation or severe air pollution events, just to name a few. The impact-based auto-labeling strategy circumvents the need for a full understanding of the relationship between the impact of interest and its driver(s). Needless to say, domain expertise is still critical in designing the auto-labeling strategy, e.g., in choosing the relevant variables and spatio-temporal scales.

\appendix
\section{Convolutional Neural Network (ConvNet)} \label{cnn}
ConvNet  has transformed image processing and pattern recognition in recent years, particularly after its performance in analyzing the ImageNet dataset in 2011 \cite{krizhevsky2012imagenet}, a turning point in the history of deep learning \cite{lecun2015deep,goodfellow2016deep}. The main components of a ConvNet algorithm are: convolutional layers, in which a specified number of filters with specified sizes extract the key features in the data and produce feature maps; Rectified Linear Unit (ReLU) layers, in which the activation function $f(x)=max(0,x)$  is applied to the feature maps to introduce nonlinearity; pooling layers, which reduce the dimensions of the feature maps by taking the maximum or average of adjacent data points, are applied to increase computing efficiency, control overfitting, and induce translational and scale invariance; and fully connected layers \cite{lecun2015deep,goodfellow2016deep}. During the training phase, patterns and their labels are inputted into ConvNet and the filters (i.e., their weights) are learned using backpropagation. 

For the small (large) domain, the Z500 patterns are on a $97\times66$ $(157\times66)$ longitude-latitude grid. In our earlier work on applying ConvNet to the LENS dataset \cite{chattopadhyay2018test}, we found the training/testing on the full-size patterns challenging. This is because the high-frequency, small-scale variability, e.g., associated with baroclinic waves, is highly chaotic and the ConvNets' attempts in learning these features are futile and lead to substantial overfitting and inaccuracies. We further found that this problem can be resolved by first down-sampling the data to $28\times28$ (using bicubic interpolation), which only retains the large-scale features. See Chattopadhyay {\it et al.}\cite{chattopadhyay2018test} for further discussions. The down-sampled patterns are then standardized by removing the mean and normalizing by the standard deviations computed over the training and testing sets combined.  
   
Our ConvNet architecture is similar to the one used in Chattopadhyay {\it et al.}\cite{chattopadhyay2018test}. There are four convolutional layers, which have $8$, $16$, $32$ and $64$ filters, respectively, and each filter has a kernel size of $5\times5$. Zero padding is used to maintain the size of patterns before and after applying the filters. Each convolutional layer is followed with a ReLU activation function. For the last two convolutional layers, the ReLU layers are followed by max-pooling layers that have a kernel size of $2\times2$ and stride of 1. The output feature is fed into a fully connected neural network with $200$ neurons. The cross-entropy cost function is accompanied by a $L_2$ regularization term, and to prevent overfitting, dropout regularization has been used in the fully connected layer. An adaptive learning rate is implemented through the ADAM optimizer \cite{kingma2014adam}. The final output is the probability of the input pattern belonging to each cluster. A softmax layer assigns the pattern to the cluster index with the highest probability.  Chattopadhyay {\it et al.}\cite{chattopadhyay2018test} provide further details on the hyperparameter choices, optimizing the ConvNet, and the effect of the number of convolutional layers (e.g., using only two layers reduces the performance while using eight leads to overfitting).   

\section{Capsule Neural Network (CapsNet)} \label{capsnet}
Despite its groundbreaking performance and ever-growing popularity, ConvNets face difficulty in dealing with some types of problems where the spatial relationship of key features become critical; e.g., a cubist painting of a face that has the position of one eye and the mouth switched has the right features of a face but not at the right relative positions. The ConvNet's challenge in dealing with these problems is due to the pooling layers, which as mentioned earlier, take the maximum or average of the adjacent data points to introduce translational and scale invariances in the deeper layers, so that ConvNet can recognize the same object at a somewhat different location or with a different size. To address this issue, Sabour {\it et al.}\cite{sabour2017dynamic} have introduced CapsNet, which does not have pooling layers, but have capsules, which are nested sets of neural layers. Unlike a neuron whose output is a real-valued scalar (e.g., likelihood of a certain feature), the output of a capsule is a vector consisting of more information (e.g., position, relative distances, poses, orientation etc.). Rather than invariance, CapsNet seeks equivariance, meaning that the relative positions of features are encoded in the feature maps. 

     One might expect CapsNet to be more suitable than ConvNet for our data-driven extreme weather prediction framework (and other climate and environmental problems), because of the importance of the relative position of features. For example, a high-pressure system on the poleward side of a low-pressure system, or a high-pressure system on the poleward side of two weaker low-pressure systems in an Omega-shape configuration, are the signature structures of persistent and extreme-causing blocking events \cite{woollings2018blocking}; however, the opposite configurations, e.g., a low-pressure system on the poleward side of a high-pressure system, often progress eastward and do not cause extreme events.        
  
     The architecture of our CapsNet is similar to that of Sabour {\it et al.}\cite{sabour2017dynamic} and is shown in Figure~\ref{fig:Clusters}. There are two convolutional layers with $32$ and $64$ filters with kernel sizes $5\times5$, followed by ReLU layers. These are followed by a primary capsule layer (with $8$ capsules, each with $8$ convolution layers) that encodes the information of the features in the form of high-dimensional tensors. Via the routing-by-agreement algorithm, the information is sent from the primary capsule layer to a secondary capsule layer, where the probability of each cluster is predicted. At each capsule, non-linearity is introduced by a squash function \cite{sabour2017dynamic} that enforces the output to have a magnitude less than one while preserving its orientation. There is also a decoding layer, a novel regularization algorithm, in which the pattern is reconstructed using $3$ fully connected layers to its original size ($28\times28$) to penalize the loss function to enforce the search for feature maps that can closely reconstruct the original pattern.    



\section{Other skill metrics: Persistence and climatology} \label{appC}
Two commonly used baseline methods for forecasting skills are climatology (wherein a variable on a particular date is forecasted as the long-term average of that variable on that particular calendar date from previous years) and persistence (where we assume that the variable remains the same over the period the forecast is performed) \cite{murphy1992climatology}. We have not compared the performance of our framework with these baseline methods because our objective could not be formulated in terms of these methods.

To predict T2m based on Z500 alone, climatology and persistence are inapplicable because they can predict Z500 patterns (or cluster) based on earlier Z500 patterns, but cannot predict T2m pattern (or cluster) based on the input of Z500 patterns, which is our objective here. To predict T2m based on Z500+T2m, persistence could not be used because by definition, there is no T2m extreme within 5 days of any temperature extreme onset. Climatology cannot be used because we have defined extremes as the 1 or 99 percentile of climatological temperature, far from the average.

\enlargethispage{20pt}

{\bf Acknowledgments:}
We thank Paige Bailey, Ashkan Borna, Packard Chan, Zhiming Kuang, Karen McKinnon, Krishna Palem, Saba Pasha, Laura Schaefer, Devika Subramanian, and Chris Wikle for insightful discussions. {We are grateful to three anonymous reviewers for helpful comments. Computational resources were provided by NSF XSEDE (allocation ATM170020) and Rice University Center for Research Computing. This work was supported by NASA grant 80NSSC17K0266 and an Early-Career Research Fellowship from the Gulf Research Program of the National Academies of Science, Engineering, and Medicine (to P.H.). A.C. thanks the Rice University Ken Kennedy Institute for Information Technology for a BP HPC Graduate Fellowship. The LENS dataset is available from \url{http://www.cesm.ucar.edu/projects/community-projects/LENS/}. The CapsNet and ConvNet codes are publicly available on GitHub: \url{https://github.com/ashesh6810/DLC_extreme}.


%
%
\bibliographystyle{unsrt} 

 \bibliography{Analog_JAMES_v1}

\end{document}